# Metal-Free Flat Lens Using Negative Refraction by Nonlinear Four-wave Mixing

**Jianjun Cao[1], Yuanlin Zheng[1], Yaming Feng[1], Xianfeng Chen[1,3], and Wenjie Wan[1,2,3*]**

[1]The State Key Laboratory of Advanced Optical Communication Systems and Networks, Department of Physics and Astronomy, Shanghai Jiao Tong University, Shanghai 200240, China

[2]University of Michigan-Shanghai Jiao Tong University Joint Institute, Shanghai Jiao Tong University, Shanghai 200240, China

[3]Key Laboratory for Laser Plasmas (Ministry of Education), Shanghai Jiao Tong University, Shanghai 200240, China

**A perfect lens with unlimited resolution has always posed a challenge to both theoretical and experimental physicists. Recent developments in optical meta-materials promise an attractive approach towards perfect lenses using negative refraction to overcome the diffraction limit, improving resolution. However, those artificially engineered meta-materials usually company by high losses from metals and are extremely difficult to fabricate. An alternative proposal using negative refraction by four-wave mixing has attracted much interests recently, though most of existing experiments still require metals and none of them has been implemented for an optical lens. Here we experimentally demonstrate a metal-free flat lens for the first time using negative refraction by degenerate four-wave mixing with a thin glass slide. We realize optical lensing effect utilizing a nonlinear refraction law, which may have potential applications in microscopy.**

Flat lenses using negative refraction create a new avenue for novel optical imaging applications, attracting intense interests from optics, microwave and even acoustic communities [1-7]. Unlike traditional optical lenses, a flat lens which can bend incoming waves at negative angles opposed to those within normal refraction regime [1-3] can form an image much more sharply thanks to its ability to negatively refract waves at all-angle including the evanescent ones, making itself a "perfect lens" to overcome the diffraction limit [2,6]. Such lenses have been realized in many formats ranging from optics, microwave to acoustic, including photonic crystals [7], metal thin film [6], meta-material [5, 9-12], etc. However, most of them suffer from high losses in association with metallic materials, which are the key elements bringing in negative permittivity and artificial permeability. Secondly, fabrications of such nano/micro structures raise additional obstacle for their practical applications. In nonlinear optics, alternative approaches to achieve negative refraction have been proposed including phase conjugation, time reversal and four wave mixing (4WM) [3,13,14]. In contrast to those artificially engineered methods i.e. meta-materials and photonic crystals, which create spatial dispersion for negative refraction using linear composition of different materials, nonlinear optics explores nonlinear wave mixings with angle matching schemes to fulfill the requirements for negative refraction. Principally, only a thin flat nonlinear slab is required. Up to now, such negative refractions using wave mixing have been realized in some thin films with high nonlinearity such as metal and graphite thin film [15-17]. However, in these experiments due to their low nonlinear conversion efficiencies or materials' optical transparency, none of them has been implemented for imaging purpose.

In this letter, we experimentally demonstrate a metal-free flat lens using negative refraction by degenerate four-wave mixing with a simple thin glass slide. Within glass slides containing third-order nonlinearity, a multi-color imaging scheme is realized at millimeter scale by converting original infrared beams into negative refracted visible ones through nonlinear wave mixings. During degenerate four-wave mixing processes, phase matching conditions enable the negative refraction of 4WM beams opposed to probe beams at some special angles. These negative refracted 4WM beams can focus and form images while slight phase mismatches due to dispersion effects or slab's thicknesses can blur images. In order to enhance the resolution of images, we study both collinear and non-collinear configurations aiming to increase numerical apertures. This new imaging technique may offer new platform for novel microscopy applications in the near future.

In a degenerate 4WM scheme, an intense pump beam at frequency $\omega_1$ and a probe beam at frequency $\omega_2$ are incident onto a slab with third order nonlinear susceptibility $\chi^{(3)}$, generating a 4WM wave at frequency $\omega_3 = 2\omega_1 - \omega_2$. Their corresponding phases should satisfy the phase matching condition: $k_3 = 2k_1 - k_2$ to ensure efficient wavelength conversion. With a thin slab material, its thickness can also affect 4WM processes, allowing 4WM with slight mismatch phases [18]. Moreover, if the slab's thickness is shorter than the wavelength, only partial phase matching is required in nonlinear surface plasmons excitations [15,19-20] or nonlinear dark-field microscope [21]. In our experiment, the pump beam at $\omega_1$ is incident on a thin glass slide normally and the probe beam at $\omega_2$ (signal beam carrying the original image) is incident at the angle of $\theta_2$ as shown in Fig. 1(a). The angle of the generated 4WM wave is denoted as $\theta_3$, measured with respect to the surface normal in counter-clockwise direction, which is negatively opposed to the normal refraction condition through a thin glass. The phase mismatch of the 4WM process reads

$$\Delta k = 2k_1 - k_2 - k_3, \quad (1)$$

where $k_i = \frac{2\pi n_i}{\lambda_i}$ $(i = 1, 2, 3)$ are the wave vectors of the pump, the probe and the 4WM beam respectively. The $n_i$ are the corresponding refractive indices of the medium. To generate an efficient 4WM wave, phase matching condition should be satisfied, i.e. $\Delta k = 0$, which leads to

$$2k_1 = k_2 \cos\theta_2^m + k_3 \cos\theta_3^m, \quad (2)$$
$$k_2 \sin\theta_2^m = -k_3 \sin\theta_3^m, \quad (3)$$

where $\theta_2^m$ and $\theta_3^m$ are the angles in the medium. These two angles are related to the angles in air by Snell's law:

$$\sin\theta_2 = n_2 \sin\theta_2^m, \quad (4)$$
$$\sin\theta_3 = n_3 \sin\theta_3^m. \quad (5)$$

Equation (3) indicates that the 4WM wave is refracted negatively with respect to the incident probe beam. Insert Eqs. (4) and (5) into Eq. (3), a Snell-like nonlinear refraction law is obtained [15]:

$$\frac{\sin\theta_2}{\sin\theta_3} = -\frac{\lambda_2}{\lambda_3}, \quad (6)$$

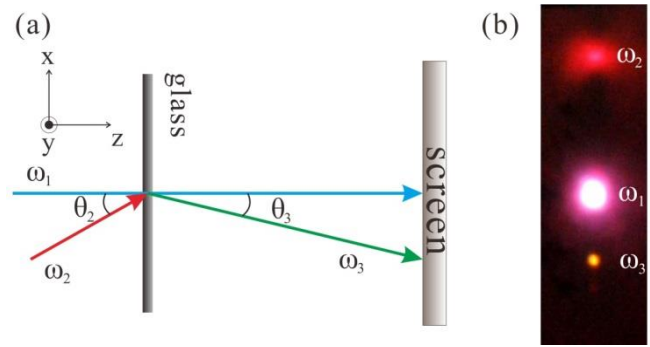

FIG. 1 (a) Schematic of negative refraction effect by degenerate 4WM in a planar glass slide. The pump beam at frequency $\omega_1$ is incident normally on the glass slide mixing with the probe beam at frequency $\omega_2$ to generate the 4WM wave at frequency $2\omega_1 - \omega_2$. The 4WM beam exhibits negative refraction opposed to the probe beam. (b) Color image on the IR viewing card. The transmitted probe beam at $\lambda_2 = 1300\ nm$ and the generated 4WM beam at $\lambda_3 = 578\ nm$ are located on the opposite side of the screen.

In our experiments, the pump beam is delivered by a Ti:Sapphire femtosecond laser source with the pulse duration of ~75 fs and central wavelength $\lambda_1 = 800$ nm. An optical parametric amplifier provides pulses of similar duration at wavelength $\lambda_2 = 1300$ nm as the probe beam. A 1 mm thick BK7 glass slide is used as the nonlinear material. The laser beams are incident on the glass slide using the geometry in Fig. 1(a), and a delay line is added to ensure the pulses' time overlapping. The generated 4WM waves after the beams passing through the glass slide are shown in Fig. 1(b). Obviously, the generated 4WM beam is negatively refracted with respect to the input probe beam, located on the other side of the central pump beam's spot. The incident angle $\theta_2$ is 7.9° and the refracted angle $\theta_3$ is $-3.4°$. By time-delaying the probe beam, the 4WM intensity varies within 165 fs time frame (see supplement), confirming that $\omega_3$ is indeed generated by a nonlinear process. The polarization of the 4WM beam is also measured to be linearly polarized in x direction the same as the input beams. However, no particular polarization is required to generate 4WM waves due to isotropic behaviors of $\chi^{(3)}$ about $2.8 \times 10^{-22} \ m^2/V^2$ in BK-7 glass[18], unlike the case of surface plasmon waves excitation by 4WM only limits to TM polarization [14,19,20]. The input powers of $\omega_1$ and $\omega_2$ are 34.5 mW and 10.3 mW. The power of the 4WM wave is 547 nW, corresponding to a conversion efficiency of $5 \times 10^{-5}$. Such beam with 547nW power is intense enough to be recorded by a visible CCD camera. The power of 4WM waves depends linearly on the probe beams' (see supplement), meanwhile the pump beams' power is limited to avoid any other nonlinear effects, e.g. self-focusing, thermal effect, that could potentially harm the nonlinear imaging quality. With the current configuration, we proceed to study the imaging formation using 4WM.

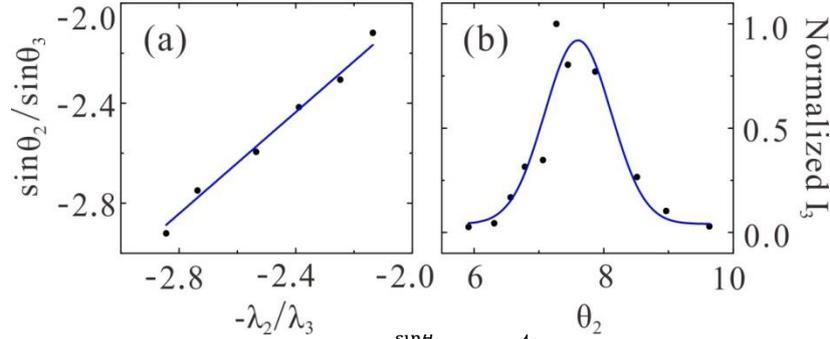

FIG. 2 (a) Nonlinear refraction law. The relationship between $\frac{\sin\theta_2}{\sin\theta_3}$ and $-\frac{\lambda_2}{\lambda_3}$ is plotted with the slope of 1.01 for the linear fitting. (b) The intensity of 4WM wave as a function of excitation angle $\theta_2$. The dots are measured 4WM intensities for probe beam at wavelength $\lambda_2 = 1300nm$. The solid curve is a Gaussian fitting.

According to the nonlinear refraction law in Eq. (6), the refracted 4WM beam's angle linearly depends on its wavelength. Our experiment measurements confirm this linear relationship between $\frac{\sin\theta_2}{\sin\theta_3}$ and $-\frac{\lambda_2}{\lambda_3}$ to have the exact slope of unit in Fig. 2(a) by varying the incident probe's wavelength from 1250nm to 1500nm while fixing the probe beam and the pump beam's angles. Such nonlinear refraction law is the direct consequence of phase matching condition in 4WM. As long as such law holds, we may extend it to a wider band of probe wavelengths accepting larger angles (numerical apertures) for imaging purpose, enabling a better imaging resolution [21,27]. However, phase matching scheme in our experiment is complicated by the ultrafast pulse's bandwidth and the glass slide's thickness. For example, though exact phase matching only allows 4WM waves to be generated at the exact angle, in real experiments, the ultrafast pulses have finite linewidth ~50nm for the pump and ~100nm for the probe, allowing 4WMs occur in a small angle spreading nearby. In Fig. 2(b), the 4WM intensity reaches its peak at $\theta_2 = 7.6°$ close to the calculated value 7.2° by Eqs. (2) and (3), while the width of the peak spreads about $\Delta\theta_2 \approx 1.0°$. Since the nonlinear refraction law in Eq. (6) depicts the linear dependence of wavelength $\lambda$ to $\sin\theta$, this angle spreading is roughly close to ~ 0.91° spreading calculated with input beams' linewidth accordingly. However, for imaging applications, such small angle spreading can cause blurry images similar to chromatic aberrations due to dispersion of a linear lens.

To experimentally realize a flat lens, we consider the phase matching condition in three-dimensional wave vector space, as shown in Fig. 3(a). The arrow ends of the incident wave vector $k_2$ that fulfill the phase matching condition in 3D compose a ring in the x-y plane. For the incident waves near the phase matching ring, their 4WM beams (green beam in Fig.3(a)) can focus on the image side according to the nonlinear refraction law. However, for a particular incidence, e.g. $k_2$ in Fig. 3(a), the phase matching ring segment is not isotropic along x-z and y-z planes, it allows a better phase matching along y-z plane rather than x-z plane in Fig. 3(b), enabling a better focus from multiple angled 4WM waves on y-z plane. In contrast, on x-z plane, only one exact phase matching angle is accepted to produce 4WM waves, giving a weak focus with poor image resolution.

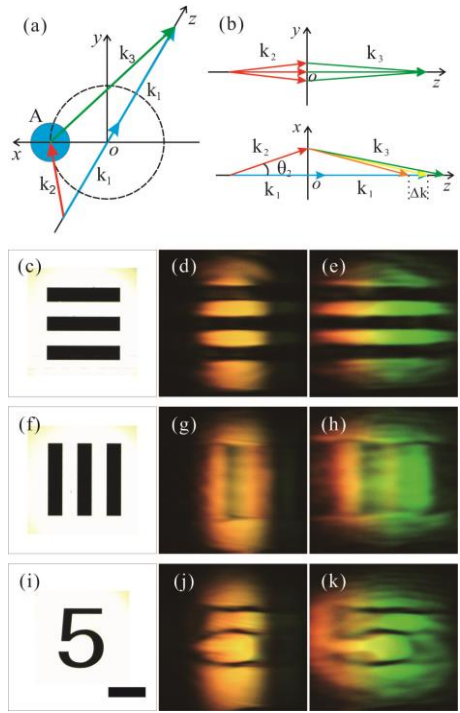

FIG. 3 Imaging a resolution card by nonlinear negative refraction effect in non-collinear configuration. (a) The phase matching condition for the degenerate 4WM process. Phase matching requires that $2k_1 - k_2 - k_3 = 0$. The dashed ring line indicates the endpoints of wave vector $k_2$ that fulfill phase matching condition. The blue disk represents the Fourier plane image of the resolution card. (b) the phase matching tringles in y-z plane and x-z plane. (c, f, i) Input images of the resolution card. (d, g, j, e, h, k) Measured images formed by 4WM wave. The thickness of the glass slide is 1 mm for (d, g, j) and 0.17 mm for (e, h, k). Scale bar, 400 μm.

First, we experimentally realize optical imaging in a non-collinear configuration as shown in Fig. 1(a), which explores the small angle cone spreading for 4WM near the phase matching ring mentioned above (see supplement). With such configuration, the probe beam with wavelength of 1300nm is incident on the glass slice at angle $\theta_2 = 7.8°$ in the x-z plane while the pump beam maintains normal incidence. A USAF resolution card is placed in the path of the probe beam, 4.3 cm away from the glass slide. 4WM waves negatively refract and form the image on the other side of the glass slide. After filtering out the probe and pump beams with optical filters, we can obtain the sharpest image with a CCD camera near the focus point at the distance of around 9.0 cm away from the glass slide. By doing this, we achieve a flat lens by probing objects with one wavelength while forming images with others. Figs. 3(c, f, i) are input images of the resolution card and Figs. 3(d, g, j) are the corresponding images formed by 4WM waves. As comparison, the image of horizontal lines is much clearer than those of vertical ones, this is because the small angle spreading caused by the object tends to have better phase matching in the y-z plane, thanks to the phase matching ring along y-z plane in Fig. 3(a), while in the x-z plane, such 4WM is less pronounced due to limited angles allowed for phase matching (the ring is a dot in x-z plane, rather than a line in y-z plane). Hence, 4WMs can be better generated and focused along y-z plane, giving a finer resolution.

To investigate this effect further, we replace the glass slide with a thinner one of 0.17 mm thickness. Clearly, color images from Fig. 3 (e, h, k) show dispersive colors ranging from red to green. Also, we perform spectroscopic measurements along the horizontal axis with a thin vertical slit in front of the fiber cable of a spectrometer (see supplement) to confirm the angle spreading of the spectrum to be around 0.6°, almost twice of that with 1mm thick glass slide. Here the thickness of thin glass slide provides an additional phase mismatching factor $\Delta k = 2\pi/d$ (d is the thickness) that allows $2k_1 - k_2 - k_3 - \Delta k = 0$, hence thinner lens leads to wider phase matching angle in x-z plane as indicated in Fig. 3(b), similar to the process in surface plasmon excitation on metal thin film [19,20]. Such phase mismatching leads to the extra angle spreading compared to the thick glass case in Fig. 3(d, g, j), resulting in the enlarged field of view in horizontal axis. Meanwhile, such wider phase matching angle also enables multicolor 4WM generation, giving rise to chromatic aberration illustrated in Fig. 3(b). Hence, image focusing in x-z plane is complicated by this phase mismatching, leaving a poor resolution. It is also worth mentioning that (1) the image size is the same as the object (see Fig. 3) as these are one-to-one correspondence with mirror symmetry, no magnification. (2) the image distance Q is related to the object distance P by $\frac{Q}{P} = -\frac{\tan\theta_2}{\tan\theta_3}$ (the segment in the x-y plane is the same).

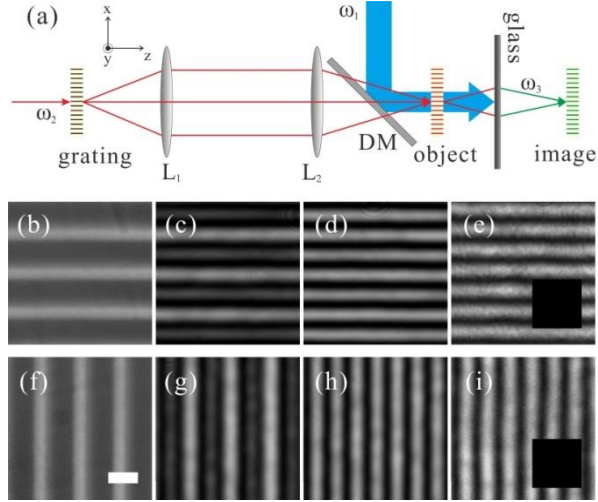

FIG. 4 Imaging a grating (object) by nonlinear negative refraction in collinear configuration. (a) Experimental setup of collinear configuration. $L_1, L_2$: lens; DM: dichroic mirror. (b, f) Images of the object. (c, g) Images of the object with grating orders $\pm 1$ and 0. (d, h) Images of the object with grating orders $\pm 1$ only by blocking 0 order, see supplement. (e, i) Measured images formed by 4WM waves. Insets show images without pump beam. Scale bar, 6.5 μm.

In order to obtain a better resolution in the transverse plane, we construct a collinear configuration to access the full phase matching ring, where pump and probe beams collinearly propagate as shown in Fig. 4(a). The pump beam at $\lambda_1 = 800\ nm$, reflected by a dichroic mirror (900 nm long pass), is incident on the glass slide normally. The probe beam at $\lambda_2 = 1350$ nm modulated by a "grating" is transformed and form an "object" (as labeled in Fig. 4) in the front of the glass slide by a 4f system using two lenses with focal lengths of 4.5cm and 6 cm respectively in order to avoid the pump beam. The image formed by the 4WM wave at $\lambda_3 = 568$ nm is recorded by a home-build microscope (see supplement). Unlike the non-collinear configuration, both vertical and horizontal lines are clear in the current configuration by taking advantage of phase matching around the full ring geometry in 3D vector space (Fig. 3). In such way, we can obtain images for both horizontal and vertical directions without the dispersion distortion as in the non-collinear configuration. However these images seem to have finer fringes (Fig. 4 (e, i)) compared to the original object in Fig. 4 (b, f), this is because that 4WM only occurs around the phase matching ring while the probe beam with normal incidence cannot efficiently generate 4WM due to phase mismatch. Here the "object" is formed by imaging the "grating", which strongly diffracts the probe beam at 0, ±1 orders. The ±1order diffractions are closer to the phase matching angles encouraging 4WM, while not for 0 order one. Hence, 4WM images (Fig. 4(e, i)) are closer to those images of the "grating" by blocking 0 order diffraction (Fig. 4 (d, h)) as opposed to those with it ( Fig. 4 (c, g)), see supplement. The resolution of the image is determined by input's numerical aperture (NA), the phase matching cone in our case, which can be estimated as $\frac{0.61\lambda_2}{\sin\theta_2} = 6$ um according to Abbe's theory [22]. Since we detect images with a shorter wavelength, ideally we could have a better resolution. Moreover, if combined with nonlinear numerical reconstruction methods in Ref. [23, 24], the resolution can be improved further with enlarged NA.

At last, we would like to comment on the direct implied applications with our flat lens: (1) Infrared (IR) microscopy, for those fluorescence dyes or biological tissues that emitting IR light, our current configuration can convert them into visible light for better detection with more sensitive visible CCD camera, e.g.

EMCCD, with a simple glass slide. However, given the low nonlinear conversion efficiency in our current experiment, flat lenses with high nonlinearity will be studied in the future. (2) compared with other flat lens by metal thin film [6] or photonic crystal [7, 25], we can achieve much larger image area up to millimeter scale determined only by pump beams size. (3) Super-resolution imaging is also possible by exploring larger angle cone of phase matching condition to accept the evanescent waves in order to break the diffraction limit, however, proper imaging reconstruction may be required; similar proposals have been proposed by using plasmonic effect in metallic nanostructures [26, 27].

**Acknowledgements**

This work was supported by the National Natural Science Foundation of China (Grant No. 11304201 and No. 61125503), the National 1000-plan Program (Youth), Shanghai Pujiang Talent Program (Grant No. 12PJ1404700).

**Correspondence**   requests for materials should be addressed to Wenjie Wan (email: wenjie.wan@sjtu.edu.cn)